\documentclass[aps,prl,twocolumn,superscriptaddress]{revtex4}
\usepackage{graphicx}
\usepackage{color}
\input epsf
\usepackage{amsmath,amssymb}
\begin{document}
\title{Acoustic analog to multiple avoided-crossings in two coupled acoustic cavities}
\author{Arjit Kant Gupta}
\affiliation{Biological Sciences and Bioengineering, Indian Institute of Technology Kanpur, Kanpur 208016, India}
\author{Anjan K. Gupta}
\affiliation{Department of Physics, Indian Institute of Technology Kanpur, Kanpur 208016, India}
\date{\today}

\begin{abstract}
A cylindrical pipe with closed ends and with a partition in-between exhibits acoustic modes in the two, thus formed, one-dimensional cavities at certain frequencies. A partial transmission through the partition leads to interaction between the two cavities' modes and to multiple avoided crossings between modes' frequencies as a function of the partition position. This is analogous to a quantum system that has two multi-level and interacting sub-systems and thus exhibits multiple avoided crossings. Such an acoustic analog is realized and studied by measuring the sound transmission as a function of frequency through a pipe with a partially transmitting and movable partition. An excellent agreement is obtained between the experimental results and a simple model based on sound wave transmission and reflection at different interfaces.
\end{abstract}
\maketitle

\section{Introduction}
When two quantum systems are made to interact with each other, their respective energy levels are modified. In particular, when the two systems have degenerate energy levels, the degeneracy is lifted. If there is an external parameter that modifies the energy levels of the systems in such a way that two systems energy levels cross each other at certain values of the external parameter, the suppression of the degeneracy manifests itself as an avoided crossing. Many physics phenomena are driven by the physics of avoided crossings. For instance, it plays a role in bird's navigation abilities \cite{av-cr-mag-birds}. It is important for quantum dynamics \cite{arul-entanglement,dehesa-shannon}. It has helped in manipulating quantum state of an atom \cite{av-cr-smm-2}, in controlling entanglement between magnetic molecules \cite{av-cr-smm-1} and in enhancing the Faraday effect \cite{nirmalaya-plasma}.

Similar behavior occurs in classical coupled oscillators or wave-media \cite{de-Anda} providing classical analogs to quantum systems or vice versa. Several classical experiments, capturing quantum analogous behavior, involving acoustic waves \cite{acc-av-cr-1,acc-band-1,acc-band-2}, em-waves \cite{av-cr-phot,ac-meta-mat} and classical oscillators \cite{av-cr-spr-mass,av-cros-class,Fano-LC}, have been reported in the recent past. These illustrate the physics of avoided crossing \cite{acc-av-cr-1,av-cr-spr-mass,av-cros-class}, Kronig-Penney like band formation \cite{acc-band-1,acc-band-2} and Fano line shape \cite{Fano-LC}. The latter refers to an asymmetric line-shape due to the interference between the waves transmitted through two different modes \cite{fano-res}. Here, one of the modes is a sharp resonance leading to a $\pi$ phase change in transmitted wave over a narrow frequency range while the other one is a broad-mode with little phase change over this frequency range.

Newman \emph{et. al.} \cite{acc-av-cr-1}, demonstrated an avoided crossing between fundamental modes of two cylindrical pipes by using a movable end-cap and a fixed partition with a hole. The allowed modes' frequencies in this experiment were found by studying frequency dependent transmission of sound which exhibits sharp resonance peaks.

In this paper, we report a systematic study of sound transmission through a pipe with a partially transmitting and movable partition. The transmittance as a function of frequency exhibits sharp peaks at certain frequencies that are described by a series of avoided crossings when the frequency of the two cavities' modes coincide. A simple model, similar to steady-state solution of Schr$\ddot{\rm o}$dinger equation in 1-D potential, or multiple reflection in thin films in optics, is discussed in detail to understand the observed frequency-dependent transmittance that arises from the underlying avoided crossings.
\section{experimental details}

We use a PVC pipe of total length $L_0=24.0$ cm, internal diameter $b=2.4$ cm and outer diameter 3.3 cm and a sheet of high density polymer foam of thickness $d=0.8$ cm to make the end-caps and the middle partition that fit tightly into the pipe. The partition can be positioned at a desired distance from one end of the pipe by pushing it with a rod. These components define the two one \begin{figure}[b]
	\centering
	\includegraphics[width=0.9\columnwidth]{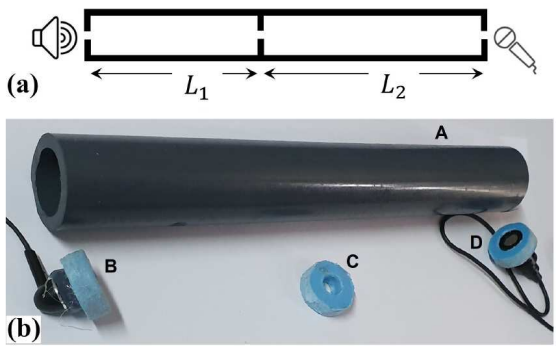}
	\caption{(a) Measurement schematic of the two acoustic cavities of length $L_1$ and $L_2$ coupled by a partition with a hole. A speaker and a receiver are fixed at the two ends for sound transmittance measurement. (b) A photograph of the actual used PVC pipe (A) of 24 cm length, end-cap with headphone speaker (B), the middle partition (C) and the and end-cap with a collar microphone receiver (D).}
	\label{fig1:exp-setup}
\end{figure}
dimensional acoustic cavities as depicted in schematic in Fig. \ref{fig1:exp-setup}(a). The end caps have machined holes to fit a small sized ear-phone speaker, commonly used with smart phones, and a collar microphone receiver. A picture of these components is shown in Fig. \ref{fig1:exp-setup}(b). A typical microphone receiver converts a sound vibration signal to an oscillating voltage which is digitized by a sound card at 440 kS/s sampling rate.  A python program \cite{suppl-info} was made to ramp the sound frequency and to measure the microphone signal's mean-squared value. The sound intensity is proportional to this mean squared amplitude signal. The frequency was ramped from 500 Hz to about 8 kHz over several tens of seconds and the mean-square-amplitude signal, averaged over a 1 Hz frequency window, was acquired as a function of frequency.

Such frequency scans were obtained over 4.5 cm to 18.5 cm distance of the partition from one end of the pipe and at 0.5 cm interval. The whole set of data was further repeated for a total of three different values of the hole diameter in the partition-disk, namely $a=0.6$, 0.8 and 1.0 cm, to study the effect of magnitude of the coupling between the two cavities.
\section{Theory}
An understanding of classical waves and their analogy to matter waves is commonly used for teaching the Schr$\ddot{\rm o}$dinger wave equation and other postulates of quantum mechanics \cite{cohen-tannudji}. In fact, the time independent Schr$\ddot{\rm o}$dinger equation for a free particle is same as the classical wave equation. This leads to identical spatial solutions for the two cases whenever the boundary conditions match for the two cases. Such a common solution then describes an eigen mode for each case with one leading to the energy of the quantum state and the other to the frequency of the classical wave. Moreover, the transmission of waves, incident on such systems, is permitted only through such resonant modes leading to sharp peaks in transmittance as a function of frequency.

\begin{figure}[t]
	\centering
	\includegraphics[width=\columnwidth]{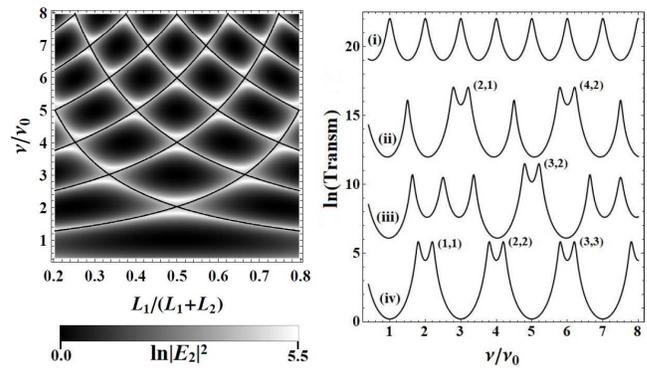}
	\caption{(a) The black intersecting lines show the variation of frequencies of different modes as a function of partition position in two cavity system with no interaction. The grey density-plot shows $\ln|E_2|^2$ with $E_2$ given by Eq. \ref{eq2:two-cav-trans} and with $r_1=r_3=0.85$, $r_2=0.8$ and $t_2=0.6i$. (b) shows the calculated $\ln|E_1|^2$ and ${\ln}|E_2|^2$ as a function of normalized frequency $\nu/\nu_0$. (i) plot is for single cavity of length $L$ (see Eq. \ref{eq1:one-cav-trans}) and for $r_1=r_3=0.85$. The other three plots, (ii), (iii) and (iv), are $\ln|E_2|^2$ with total length $L$ and with $L_1/L=1/2,3/5$ and 2/3, respectively. $(n,m)$ indicate the otherwise degenerate modes of the two cavities that lead to the marked split resonance peak due to coupling. These plots have been shifted vertically for clarity.}
	\label{fig2:freq-cross}
\end{figure}

The acoustic modes in a closed cavity of length $L$ are governed by the existence of antinodes in pressure field, or nodes in the displacement field, at the two closed ends \cite{kinsler-book}. This leads to permitted wavelengths $\lambda_n=2L/n$ and frequencies $\nu_n=n\nu_0$ with $\nu_0=v_s/2L$ and $v_s$ as sound-speed. For a pipe of length $L$ having a solid partition at distance $L_1$ from one end, one gets two decoupled or non-interacting cavities with each having acoustic modes independent of the other. The frequency of these modes is given by $\nu_{1n}=nv_s/2L_1=n\nu_0/x$ and $\nu_{2m}=mv_s/2L_2=n\nu_0/(1-x)$. Here, $x=L_1/(L_1+L_2)$. As shown in Fig. \ref{fig2:freq-cross}(a) by black lines, the frequencies of the two sides' modes exhibit multiple crossings as a function of the partition position. For instance, for symmetric partition position, i.e. $x=0.5$, the two cavities have identical mode frequencies $2n\nu_0$. In general for a given rational $x=n/(n+m)$, the frequencies of the $n^{\rm th}$ mode of one cavity and $m^{\rm th}$ mode of the other, together with their higher harmonics, will coincide. Here, $m$ and $n$ are non-zero positive integers. This is the case when the partition is solid and thus it does not allow any interaction between the waves of the two cavities. In case the partition allows some transmission these frequency crossings become avoided crossings as discussed further.

We first discuss transmission through a single 1-D cavity using two different methods. The first method consists of keeping track of all the reflected and transmitted waves at the two interfaces. This is similar to the derivation used in wave-optics for anti-reflection coatings \cite{Hecht-optics-book}. The second method is similar to that used in quantum mechanics \cite{cohen-tannudji} where one works out the position dependent wave-function by solving the time-independent Schr${\ddot{\rm o}}$dinger equation for a given energy. One can thus find the transmittance and reflectance involving simple or more involved 1-D potentials \cite{delta-array}. Here we solve for the time-independent wave-equation for a given frequency $\nu=v_sk/2\pi$. Here, $v_s$ is the sound velocity and $k=2\pi/\lambda$ is the wave-vector. A plane wave $e^{ikx}$, incident on a cavity with two partially reflecting ends (referred to as called interfaces), undergoes transmission and reflection at first interface. The wave thus transmitted undergoes reflection and transmission at the second interface and this reflected wave strikes back at the first interface to again get reflected and transmitted and so on.

\begin{figure}[h!]
	\centering
	\includegraphics[width=0.75\columnwidth]{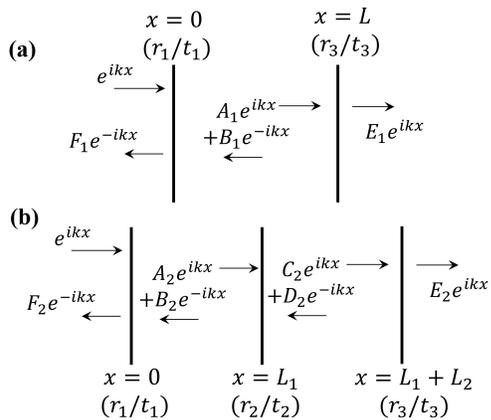}
	\caption{Schematic for steady-state pressure-wave configuration for one-dimensional waves for finding the net transmittance: (a) for single cavity of length $L$ and, (b) for two cavities of lengths $L_1$ and $L_2$ with a partially transmitting interface in-between.}
	\label{fig3:model-schem}
\end{figure}

We define $r_i$ and $t_i$ as the complex amplitude-transmission and amplitude-reflection coefficients, respectively, for sound incidence direction from inside to outside the cavity. Here, $i=1,3$ for the two cavity ends, see Fig. \ref{fig3:model-schem}(a). These satisfy $|r_i|^2+|t_i|^2=1$, assuming no energy loss at the interfaces. $r_i'$ and $t_i'$ refer to the sound incident in opposite direction as compared to the un-primed quantities. In general, $r_i'\neq r_i$ or $t_i'\neq t_i$. This way the amplitude of the net transmitted wave, after multiple reflections in the cavity, works out as  $t_1't_3(1+r_1r_3p^2+r_1^2r_3^2p^4+...)$, i.e. $t_1't_3/(1-r_1r_3p^2)$. Here $p=e^{ikL}$, with $k=2\pi\nu/v_s$, describe the phase change during the traversal of the wave from $x=0$ to $L$.

An alternative method uses the steady-state wave configuration, as depicted in Fig. \ref{fig3:model-schem}(a), with net $A_1e^{ikx}+B_1e^{-ikx}$ between the two interfaces and the net reflected and transmitted waves as $F_1e^{-ikx}$ and $E_1e^{ikx}$, respectively. By analyzing the transmission and reflection at the two interfaces, at $x=0$ and $x=L$, we get four linear equations, namely, $r_1'+t_1B_1=F_1$, $t_1'+r_1B_1=A_1$, $t_3e^{ikL}A_1=E_1e^{ikL}$ and $r_3e^{ikL}A_1=e^{-ikL}B_1$. Solving these leads to the same outcome, i.e.
\begin{equation}
E_1=\frac{t_1't_3}{1-r_1r_3p^2}.
\label{eq1:one-cav-trans}
\end{equation} This shows that for $r_1, r_3\approx 1$, $E_1$ will have sharp peaks for $p^2=e^{2ikL}=1$, i.e. $kL=n\pi$. Thus the overall transmittance, i.e. $|E_1|^2$, will exhibit peaks at frequencies $\nu=n\nu_0$, as expected and shown in the plot (i) in Fig. \ref{fig2:freq-cross}(b). Note that the only parameter that affects this plot is product $r_1 r_3$. The height of the peaks diverges when $|r_1 r_3|=1$ and it decreases when $|r_1 r_3|$ is reduced. The phase of $r_1 r_3$ leads to a uniform shift in all the peak-frequencies. The second method is more tractable, particularly when there are multiple interfaces as discussed further.

For two cavities separated by a partially transmitting interface, we look at the steady-state configuration of waves as shown in the schematic in Fig. \ref{fig3:model-schem}(b). We define $p_1=e^{ikL_1}$ and $p_2=e^{ikL_2}$ and analyze the transmission and reflection at the three interfaces, having transmission and reflection coefficients as $t_i$ and $r_i$ ($i=1,2,3$), respectively. This leads to the following six equations,
\begin{align*}
t_1'+r_1B_2&=A_2\\
r_1'+t_1B_2&=F_2\\
r_2p_1A_2+t_2p_1^{-1}D_2&=p_1^{-1}B_2\\
t_2p_1A_2+r_2p_1^{-1}D_2&=p_1C_2\\
r_3p_1p_2C_2&=p_1^{-1}p_2^{-1}D_2\\
t_3p_1p_2C_2&=p_1p_2E_2\\
\end{align*}
For instance, the third equation is obtained by analyzing $B_2e^{-ikx}$ wave as resulting from the superposition of the waves reflected from $A_2e^{ikx}$ and transmitted from $D_2e^{-ikx}$ at the second interface. One has to keep track of the phase factors associated with each wave. Also for a partition-disk with a uniform diameter hole: $r_2'=r_2$ and $t_2'=t_2$. The second approach here is similar to the transfer matrix method \cite{tr-matrix-book} for waves in terms of getting the above linear equations. These linear equations can be solved for $A_2,B_2,C_2,D_2,E_2$ and $F_2$. The net transmittance will be given by $|E_2|^2$ and $E_2$ works out as,
\begin{equation}
E_2=\frac{t_1't_2t_3}{(1-p_1^2r_1r_2)(1-p_2^2r_2r_3)-p_1^2p_2^2r_1r_3t_2^2}.
\label{eq2:two-cav-trans}
\end{equation}
This is consistent with the earlier single cavity expression, Eq.\ref{eq1:one-cav-trans}, in the limit of no middle interface i.e. $r_2=0$ and $t_2=1$, after realizing that $p_1^2p_2^2=p^2$.

The resulting transmission is shown as a two-dimensional density-plot of $\ln|E_2|^2$ in Fig. \ref{fig2:freq-cross}(a) as a function of $x=L_1/(L_1+L_2)$ and $\nu/\nu_0=v_sk/2\pi\nu_0$ with $\nu_0=v_s/2L$. We have plotted this for frequency independent $r_1=r_3=0.85$, $r_2=0.8$ and $t_2=0.6i$. The same figure also depicts the mode-frequencies, as black lines, for zero coupling to depict the effect of avoided crossings in the transmittance for non-zero coupling. The plots (ii)-(iv) in Fig. \ref{fig2:freq-cross}(b) show $\ln|E_2|^2$ for $x$ = 1/2, 3/5 and 2/3, respectively. For $x=1/2$ all the modes of the two cavities are degenerate leading to a split peak for each mode frequency. For $x=3/5$, the third mode of one cavity and second mode of the other, labeled as (3,2), and their higher harmonics, are degenerate which show split peaks while other non-degenerate modes show a single peak. Similarly for $x=2/3$ one can see that modes (2,1) and higher harmonics are degenerate.

A phase difference of nearly $\pi/2$ seems to be necessary between $t_2$ and $r_2$ in order to have the avoided crossings in qualitative agreement with experiments discussed later. If one plots $E_2$ Vs frequency for fixed magnitude of $r_2$ and $t_2$ but with varying phase difference, the splitting is maximum with symmetric peaks at $\pi/2$ phase difference as seen in Fig. \ref{fig2:freq-cross}(b). As this phase-difference is changed from $\pi/2$ the two peaks become unequal and at zero or $\pi$ phase difference one of the two splitted peaks disappears. In fact $r_2$ and $t_2$ are dependent on the pipe's diameter and on the thickness of the disk and diameter of the hole in it. This is discussed in Appendix. The frequency dependent phase-difference between $r_2$ and $t_2$ is indeed found to be close to $\pi/2$, particularly at low frequencies.

It can be easily seen that the phases of $t_1'$ and $t_3$ do not matter for the variation of $|E_2|^2$ with frequency but the details of complex $r_i$ and $t_2$ and their frequency dependence are important for a comprehensive understanding of the net transmittance. The exact phase and frequency dependence of $r_1$ and $r_3$ will depend on the details of the speaker and the microphone receiver. However, in the limit of small openings, in the end caps, as well as the partition, we expect the phases of all $r_i$ to be small in magnitude. These phases are also expected to be negative as the reflected waves should leg behind the incident wave.

\begin{figure}[h!]
	\centering
	\includegraphics[width=0.9\columnwidth]{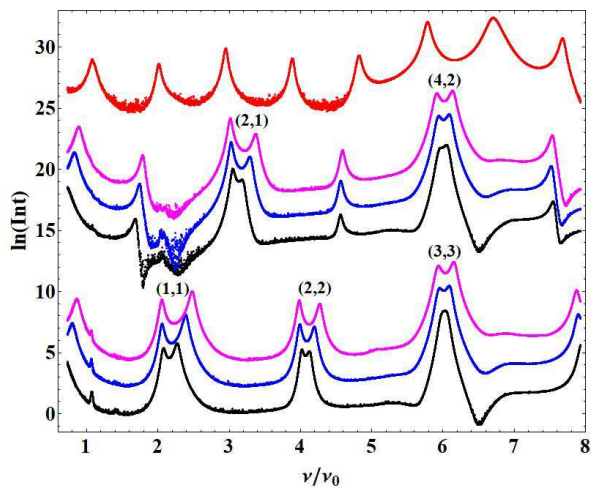}
	\caption{Measured sound intensity as a function of frequency. The top red curve is for no partition. The bottom three curves correspond to $x\approx 1/2$ with the black line for partition hole diameter $a=0.6$ cm, blue line for $a=0.8$ cm and magenta line for $a=1.0$ cm. The middle three curves are for a partition at 3.5 cm from the pipe center which corresponds to $x\approx2/3$ with respective colors corresponding to the same three $a$ values. Note that graphs have been offset vertically for clarity.}
	\label{fig4:transm-spectra}
\end{figure}
\section{results and discussion}
Fig. \ref{fig4:transm-spectra} shows the measured frequency dependent transmitted sound intensity for the pipe without partition and with the partition at two different positions for three different sizes of the hole in the partition. The top red plot, for the pipe without partition, shows sharp peaks expected at frequencies $\nu_n=n\nu_0$. Using this plot we obtain $\nu_0=v_s/2L=0.79(\pm0.02)$ kHz from the difference between successive peaks. With effective cavity length as $L=L_0-2d=22.4$ cm the value of the sound speed works out as $v_s=354(\pm 9)$ m/s. The accepted value of sound at 20$^\circ$C is about 343 m/s and it has $\sqrt{T}$ dependence on absolute temperature $T$. We did not keep track of the precise room temperature during the measurement. We expect it to be between 25 and 35$^\circ$C.

The bottom three plots in Fig. \ref{fig4:transm-spectra} show the transmitted intensity for partition-disk having $a=0.6$, 0.8 and 1.0 cm diameter holes and with partition-disk centered at the pipe-center. We clearly see doubly split peaks with each pair centered close to frequency $nv_s/L'$ with $2L'=L_0-3d$. Due to the reduction in $L_1+L_2$ by $d$ from the presence of the partition, the effective value of $\nu_0$ for this case is $v_s/2L'=0.82$ kHz, which is slightly higher than that without the partition. This leads to a slight mismatch between single cavity, i.e. red plot, and the coupled cavity plots. Thus, for instance, there is a mismatch between the second peak of the red plot and the center of the (1,1) split peak in the bottom black-plot. As expected, the splitting magnitude is seen to grow with increasing hole-diameter in the partition.

\begin{figure}[h!]
	\centering
	\includegraphics[width=0.99\columnwidth]{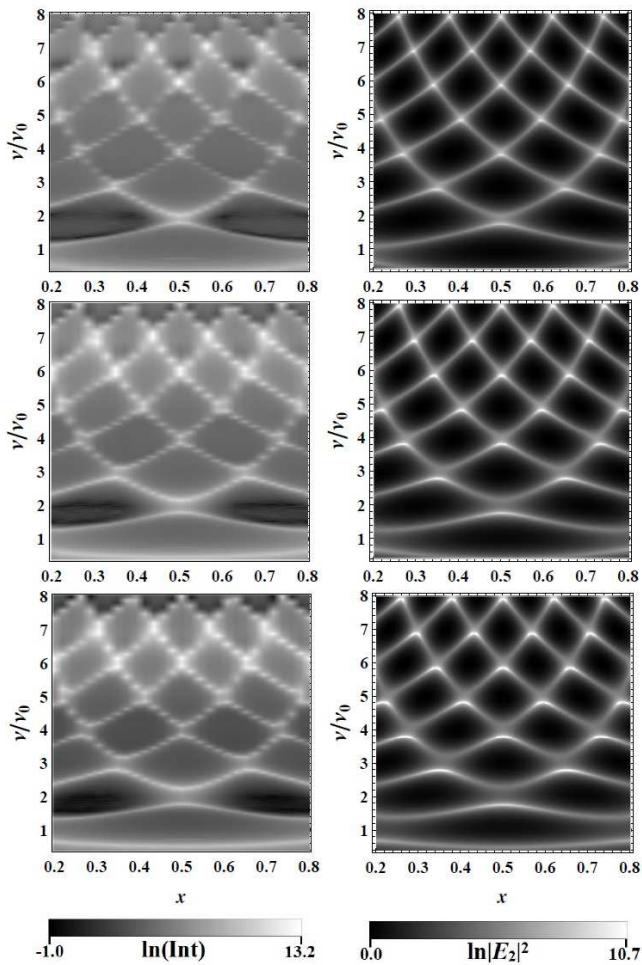}
	\caption{The left column shows the density-plots of the measured transmitted intensity (on ln scale) as a function of partition position and normalized frequency $\nu/\nu_0$, with $\nu_0=0.82$ kHz, for three different partition hole diameters, i.e. $a=0.6, 0.8, 1.0$ cm, from top to bottom, respectively. The right column shows the corresponding calculated transmitted-intensity density-plots, i.e. $\ln|E_2|^2$, with $r_1=r_3=0.85e^{-0.3\pi i}$ and $r_2$ \& $t_2$ given by Eqs. \ref{eq4:part-refl} \& \ref{eq3:part-transm}. There is a vertical offset between different plots for clarity}
	\label{fig5:density-plots}
\end{figure}
The two sides's modes are not degenerate for a general partition position. However, as discussed earlier the degeneracy occurs for certain partition positions and between different modes of the two sides. The middle three plots in Fig. \ref{fig4:transm-spectra} show how the peaks change when the partition is moved away from the center of the pipe by 3.5 cm distance leading to $x$ close to 2/3. The avoided crossing thus occurs between the 1$^{\rm st}$ and 2$^{\rm nd}$ modes and between their higher harmonics. A general sound transmission curve as a function of frequency is not easy to comprehend for an arbitrary partition position. However, it becomes remarkably clear when one looks at the two dimensional density-plot of the sound intensity on log-scale as a function of partition-position and frequency. We see a good qualitative agreement between the experimental density-plots, shown in left column of Fig. \ref{fig5:density-plots}, and theoretical one in Fig. \ref{fig2:freq-cross}(a). These also show a growth in frequency difference at each avoided crossing with an increase in the partition-hole size.

For a given hole-size the frequency difference at avoided crossings decreases with increasing frequency. This is not captured by the frequency-independent $r_2$ and $t_2$. However, with a more realistic model for the sound transmission and reflection at the partition disk with a hole, as discussed in Appendix, one gets a very good agreement between the model and the experiment as can be seen by comparing the density-plots in Fig. \ref{fig5:density-plots}. The plots in the right column use the frequency dependent $r_2$ and $t_2$, as discussed in Appendix, with actual $S_1$ and $S_2$ values. This is now clearly understandable from the variation with frequency of the phase-difference between $r_2$ and $t_2$ and their magnitudes. As seen in Fig. \ref{fig6:appendix}(c), their phase difference decreases with frequency while the magnitude of $t_2$ reduces leading to reduced coupling between the two cavity's modes with increasing frequency.

Though a very good agreement is seen between the experiment and the model, there are a few points that need discussion. A peak is also seen in the two cavity data in 500-800 Hz range that rises to higher frequencies as the partition hole diameter is increased, see Fig. \ref{fig4:transm-spectra} and \ref{fig5:density-plots}. This occurs as the avoided crossing corresponding to the zero frequency single-cavity mode leads to two low frequency peaks. Our experimental data were quite noisy below about 200 Hz making it difficult to see the lowest frequency peak. Additionally, all the resonance peaks shift to higher frequencies when one incorporates a negative phase in both $r_1$ and $r_3$. The best agreement with experiment is obtained for $-0.3\pi$ phase for both $r_1$ and $r_3$. Unequal phases of $r_1$ and $r_3$ lead to a shift of the density-plot either towards lower $x$ or higher so that the symmetric point shifts away from $x=1/2$. The effect of such a phase can also be seen in the single cavity data in Fig. \ref{fig4:transm-spectra} as the lowest frequency peak occurs close to 900 Hz which is slightly above $\nu_0$.

There are two minor drawbacks in our experimental setup that rule out the quantitative comparison of the measured intensity with the model. First, the combined response of the speaker and microphone has a frequency dependence leading to a broad peak centered close to 5.5 kHz frequency. This can clearly be seen from the single cavity red-plot in Fig. \ref{fig4:transm-spectra}. Second, some sound gets transmitted from speaker to microphone through media, including the cylinder walls, other than the intended cavity. Although this is very low in intensity but it affects the shape of some of the resonance peaks due to interference between the sounds reaching through the two pathways. The effect is more visible when the sound intensity through the cavity becomes comparable to the other media. As a result of destructive interference, just after some of the peaks, one can see the sound intensity dipping below the background level and the peaks acquiring a Fano-like asymmetry \cite{fano-res}. In fact this effect was seen to be more drastic with another microphone which was not so snugly fitted into the end cap leading to more sound transmission due to media other than cavity.

\section{conclusions}
In conclusion, our inexpensive homemade setup illustrates a rather complex, though systematic, series of avoided crossings in resonance frequencies of two coupled one-dimensional acoustic cavities. This is described by a theoretical model using reflection and transmission at different interfaces, which is accessible to senior undergraduates and quite generic to be useful for different waves. The model uses the actual experimental geometry with an excellent agreement with results. This experiment can be easily setup in an undergraduate laboratory to illustrate multiple avoided crossings with a simple and rather comprehensive theoretical understanding.

\section{appendix}
\begin{figure}[b]
	\centering
	\includegraphics[width=1.0\columnwidth]{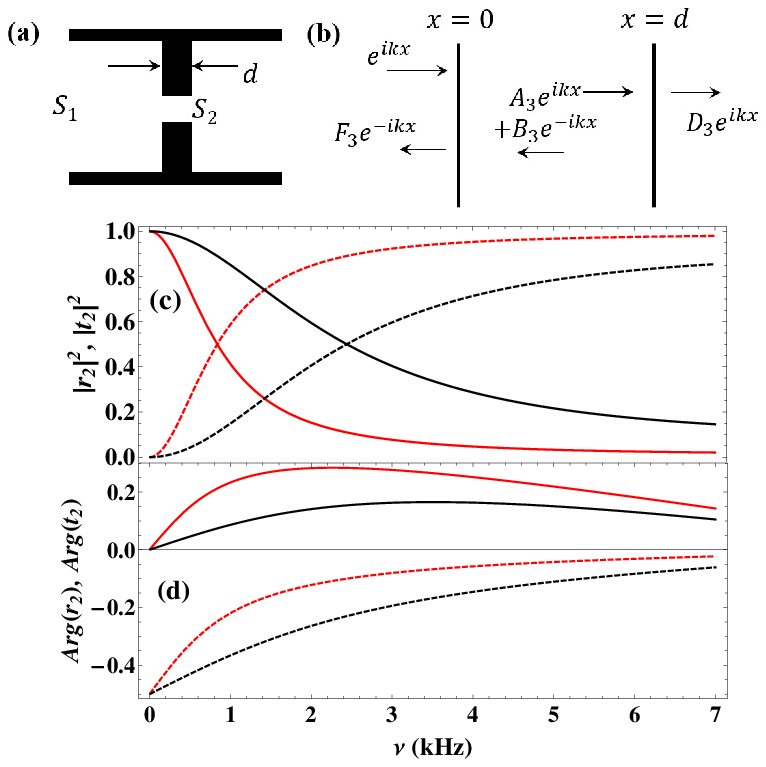}
	\caption{Transmission through a disk of thickness $d$ having a hole of cross-section area $S_2$ placed inside a circular pipe of cross-section area $S_1$. (a) shows the schematic and (b) depicts the steady-state pressure-wave configuration for a unit amplitude incident wave leading to the resulting reflected and transmitted waves together with the waves in the disk-region. (c) shows the magnitude of $r_2$ (broken lines) and $t_2$ (continuous lines) for $S_2/S_1=1/16$ (red) and for $1/9$ (black). (d) shows the corresponding phases in $\pi$ units. Note that (c) and (d) plots use $\nu$ in kHz unit with $L=24$ cm and $d=0.8$ cm.}
	\label{fig6:appendix}
\end{figure}
Fig. \ref{fig6:appendix}(a) shows the schematic of a pipe of cross-section area $S_1=\pi b^2/4$ with a partition disk of thickness $d$ having a hole of cross-section area $S_2=\pi a^2/4$. A sound wave of unit amplitude is incident from left resulting into a net reflection and transmission due to impedance change at the two planes defined by the disk thickness $d$. From the continuity of pressure and volume velocity at the $ij$-interface one gets $r_{ij}=(S_i-S_j)/(S_i+S_j)$ and $t_{ij}=2S_i/(S_i+S_j)$ \cite{kinsler-book}. Note that $r_{ij}=-r_{ji}$. One gets four equations, namely $r_{12}+t_{21}B_3=F_3$, $t_{12}+r_{21}B_3=A_3$, $t_{23}A_3=D_3$ and $r_{23}A_3e^{2ikd}=B_3$. This leads to an overall amplitude reflection coefficient across the disk as $r_2=F_3=(r_{12}+r_{23}e^{2ikd})/(1+r_{12}r_{23}e^{2ikd})$ and $t_2=D_3=t_{12}t_{23}/(1+r_{12}r_{23}e^{2ikd})$. Defining $r_0=(S_1-S_2)/(S_1+S_2)=r_{12}=-r_{21}=r_{32}=-r_{23}$ and using $t_{12}t_{23}=1-r_0^2$, this simplifies to
\begin{align}
r_2&=\frac{-2r_0 i\sin kd}{(1-r_0^2)\cos kd-i(1+r_0^2)\sin kd}
\label{eq4:part-refl}\\
t_2&=\frac{(1-r_0^2)e^{-ikd}}{(1-r_0^2)\cos kd-i(1+r_0^2)\sin kd}.
\label{eq3:part-transm}
\end{align}
Note that these expressions satisfy the expected relation $|r_2|^2+|t_2|^2=1$ and for this symmetric configuration the above expressions will be same for sound incident from either side of the disk. Moreover, for $kd<<1$, $r_2$ and $t_2$ have a phase difference of $\pi/2$. This phase difference reduces from $\pi/2$ and $|t_2|$ decreases from unity value with increasing frequency as seen in Fig. \ref{fig6:appendix}(c) and (d).

\end{document}